\documentclass[10pt,twocolumn]{article}

\usepackage{abstract}
\usepackage{amsmath}
\usepackage[textwidth=16.6cm, textheight=22.9cm]{geometry}
\usepackage{graphicx}
\usepackage{multirow}
\usepackage{subfig}

\begin{document}

\pagenumbering{arabic}
\title{\sf\textbf {A Performance Model for the Communication in Fast Multipole Methods on HPC Platforms}}
\author{\sf\textbf {Huda Ibeid, Rio Yokota, and David Keyes}\\
\sf\normalsize Division of Computer, Electrical and Mathematical Sciences and Engineering\\
\sf\normalsize King Abdullah University of Science and Technology, Saudi Arabia}
\date{}

\twocolumn[
\maketitle
\begin{onecolabstract}
Exascale systems are predicted to have approximately one billion cores, assuming Gigahertz cores. Limitations on affordable network topologies for distributed memory systems of such massive scale bring new challenges to the current parallel programing model. Currently, there are many efforts to evaluate the hardware and software bottlenecks of exascale designs. There is therefore an urgent need to model application performance and to understand what changes need to be made to ensure extrapolated scalability. The fast multipole method (FMM) was originally developed for accelerating $N$-body problems in astrophysics and molecular dynamics, but has recently been extended to a wider range of problems, including preconditioners for sparse linear solvers. It's high arithmetic intensity combined with its linear complexity and asynchronous communication patterns makes it a promising algorithm for exascale systems. In this paper, we discuss the challenges for FMM on current parallel computers and future exascale architectures, with a focus on inter-node communication. We develop a performance model that considers the communication patterns of the FMM, and observe a good match between our model and the actual communication time, when latency, bandwidth, network topology, and multi-core penalties are all taken into account. To our knowledge, this is the first formal characterization of inter-node communication in FMM, which validates the model against actual measurements of communication time.
\vspace{3em}
\end{onecolabstract}
]

\section{Introduction}
$N$-body problems arise in many areas of physics (e.g. astrophysics, molecular dynamics, acoustics, electrostatics). In these problems, the system is described by a set of $N$ particles and the dynamics of the system arise from interactions that occur between every pair of particles. This requires $\mathcal{O}(N^2)$ computational complexity. For this reason, many efforts have been directed at producing \textit{fast} $N$-body algorithms. More efficient algorithms of the particle interaction problem can be provided by a hierarchical approach using tree structures. In this approach, the computational domain is hierarchy subdivided, and the particles are clustered into a hierarchical tree structure. The approximation is applied to far-field interactions, whereas near-field interactions are summed directly. When the far-field expansion is calculated against the particles directly, this approach called a treecode \cite{Barnes1986}. When the far-field effect is translated to local-expansions before summing their effect, it is called a fast multipole method (FMM) \cite{Greengard1987, Cheng1999}. These approaches bring the complexity down to $\mathcal{O}(N\log N)$ and $\mathcal{O}(N)$ for treecode and FMM, respectively. FMM has been listed as one of the top ten algorithms of the twentieth century \cite{Dongarra2000} due to its wide applicability and impact on scientific computing. It was originally developed for applications in electrostatics and astrophysics, but continues to find new areas of application such as aeroacoustics \cite{Wolf2011}, fluid dynamics \cite{Greengard1996a}, magnetostatics \cite{VandeWiele2008}, and electrodynamics \cite{Zhao2000}. Because of its linear complexity, FMMs scale well with respect to the problem size, if implemented efficiently.

Since the performance of  a single-processor core has plateaued, future supercomputing performance will depend mainly on increases in system scale rather than improvements in single-processor performance. Processor counts are now going from hundreds of thousands to millions, which means that the number of cores, interconnect, and memory will grow enormously. For this reason, modeling application performance at these scales and understanding what changes need to be made to ensure continued scalability on future exascale architectures is necessary. Since the performance of the FMM has a large impact on a wide variety of applications across a wide range of disciplines, it is important to understand the challenges that FMMs face on future architectures with increased parallelism, as well as to predict and locate bottlenecks that might cause performance degradation.

The present study develops and demonstrates a performance model for the communication in FMM. To model the performance, we start with the baseline model which is the basic $\alpha-\beta$ model for communication, where $\alpha$ is the latency and $\beta$ is the inverse bandwidth. Then, some penalties are added to the baseline model based on machine constraints. These penalties include distance and reduced per-core bandwidth. We validate our performance model on three different architectures, Shaheen (BG/P), Mira (BG/Q), and Titan (Cray XK7).

The paper is organized as follows. Section 2 gives an overview of related work. Section 3 summarizes some performance challenges that face FMM on parallel machines. These challenges include massive parallelism and degradation due to inter-node communication. In Section 4, an exposition of the fast multipole method sufficiently detailed to expose communication properties is given. Section 5 describes our performance model. Experiments done to validate the performance models are provided in Section 6. Lessons from the model results are presented in Section 7 and we conclude in Section 8.

\section{Related work}
Performance modeling and characterization for understanding and predicting the performance of scientific applications on HPC platforms has been targeted by many related projects. For example, Clement and Quinn developed a performance prediction methodology through symbolic analysis of their source code \cite{Clement1995}. Mendes and Reed focused on predicting scalability of an application program executing on a given parallel system \cite{Mendes1998}. Mendes proposed methodology to predict the performance scalability of data parallel applications on multi-computers based on information collected at compile time \cite{Mendes1997}. The approach of combining computation and communication to obtain a general performance model is described by Snavely \textit{et al.} \cite{Snavely2001}. DeRose and Reed concentrate on tool development for performance analysis \cite{DeRose1999}. Performance models for a specific given application domain, which presents performance bounds for implicit CFD codes have also been considered \cite{Gropp1999}. The efficiency of the spectral transform method on parallel computers has been evaluated by Foster \cite{Foster1997}. Kerbyson \textit{et al.} provide an analytical model for the application SAGE \cite{Kerbyson2001}. Performance models for AMG were developed by Gahvari \textit{et al.} \cite{Gahvari2011}. Traditional evaluation of specific machines via benchmarking is presented by Worley \cite{Worley2000}.

Scaling FMM to higher and higher processors counts has been a popular topic \cite{Perez-Jorda1998, Jetley2010}, while extensive study of single-node performance optimization, tuning, and analysis of FMM has also been of interest \cite{Vuduc2010}. However, there has been little effort to model the inter-node communication of FMMs. Lashuk \textit{et al.} derive the overall complexity of FMM on distributed memory heterogeneous architectures \cite{Lashuk2009}, but do not validate the model against the actual performance. The present work is based on the communication model for AMG \cite{Gahvari2011}, and extends their theory to FMM. To our knowledge, this is the first formal characterization of inter-node communication in FMM, which validates the model against actual measurements of communication time.

\section{Performance challenges}
High performance computing systems have shown a fast and sustained growth with performance improvement of 10x every 3.6 years. This performance improvement comes at a high cost and introduces many challenges. Furthermore, the development of an exascale computing capability will cause significant and dramatic changes in computing hardware architecture relative to current petascale computers. In this section we present some of the challenges faced by FMMs to achieve good parallel performance on future exascale systems. 

\subsection{Trends in Computer Hardware}
Massively parallel machines have emerged as the most widely used high-performance computing platforms. These machines are characterized currently by hundreds of thousands of computing nodes, and this number will continue to grow.  Another trend of massively parallel machines is to increase the numbers of cores on each node. These nodes communicate by sending messages through a network, which leads to lower scalability and less performance due to cores on a single node contenting for access to the interconnect. We discuss multicore issues in more detail when presenting our performance models that take this into account.

\subsection{Communication}
Algorithms have two costs in terms of time and energy: computation (flops) and communication (Bytes). Communication involves moving data between levels of a memory hierarchy in case of sequential algorithms and exchanging data between processors over a network in the case of parallel algorithms. Therefore, without considering overlap, the running time of an algorithm is the sum of three terms: the number of flops times the time per flop, the number of words moved divided by the bandwidth (measured as words per unit time), and the number of messages times the latency. The last two terms are time consumed by communication. Given that the time per flop is generally much less than the reciprocal of bandwidth for most applications, we can see that inter-node communication eventually consumes most of the running time for large scale calculations. Furthermore, the gaps between computation and communication are growing exponentially with time, as we can see in Table \ref{tab:gap}.

Communication latency and bandwidth are becoming bottlenecks, where the latency is governed by physics and the bandwidth by cost. Therefore, minimizing communication within the algorithm becomes crucial. Appropriate performance models can guide development of algorithms to help reduce the communication.

\begin{table}[!t]
\renewcommand{\arraystretch}{1.3}
\caption{Differential rates of improvement of computation and communication}
\label{tab:gap}
\centering
\begin{tabular}{|c||c||c||c|}
\hline
 & \multicolumn{3}{c|}{Annual Improvements}\\
\hline
\(flop rate\) & & \(bandwidth\) & \(latency\)\\
\hline
\multirow{2}{*}{\(59\%\)} & Network & \(26\%\) & \(15\%\)\\
& DRAM & \(23\%\) & \(5\%\)\\
\hline
\end{tabular}
\end{table}

\begin{figure*}[!t]
\centerline{\subfloat[2-D view]{\includegraphics[width=0.25\textwidth]{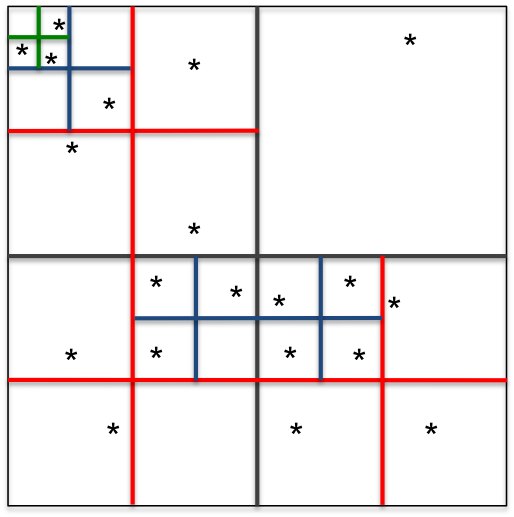}%
\label{2-D view}}
\hfil
\subfloat[Tree view]{\includegraphics[width=0.4\textwidth]{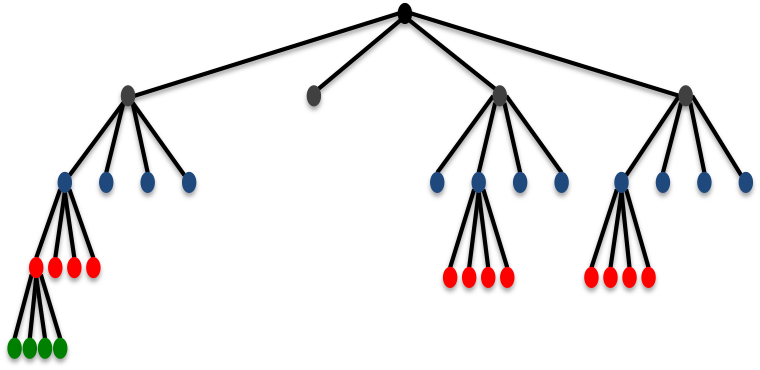}%
\label{Tree view}}}
\caption{Hierarchical decomposition}
\label{fig:tree}
\end{figure*}

\section{Fast multipole method}
$N$-body methods are most commonly used to simulate the interaction of particles in a potential field, which has the form
\begin{equation}
f(\mathbf{x}_i) = \sum\limits_{j=1}^N q_jK(\mathbf{x}_i,\mathbf{x}_j)
\end{equation}

Here, $f(\mathbf{x}_i)$ represents a field value evaluated at a point $\mathbf{x}_i$ which is generated by the influence of sources located at $\mathbf{x}_j$ with weights $q_j$. $K(\mathbf{x}_i,\mathbf{x}_j)$ is the kernel that governs the interactions between evaluation and source particles. The direct approach to simulate the $N$-body problem is relatively simple; it evaluates all pair-wise interactions among the particles. While this method is exact to within machine precision, the solution is $\mathcal{O}(N^{2})$ in its computational complexity, which is prohibitively expensive for even modestly large data sets. However, its simplicity and ease of implementation make it an appropriate choice when simulating small particle sets $(N<100)$ where high accuracy is desired \cite{Rankin1999}. For a larger number of particles, many faster algorithms have been invented, e.g. treecode and fast multipole method (FMM). The main idea behind these fast algorithms is to approximate the effect of sufficiently far particles. The most common way to achieve this approximation is to cluster the far particles into larger and larger groups by constructing a tree structure. The treecode clusters the far particles and achieves $\mathcal{O}(N \log N)$ complexity. The FMM further clusters the near particles in addition to the far particles to achieve $\mathcal{O}(N)$ complexity.

In this section, we present an overview of fast algorithms that have been developed for the calculation of $N$-body problems. First, the spatial hierarchy and the fast approximate evaluation of these algorithms are discussed. Then, a description of the communication introduced by the domain partitioning scheme used in these algorithms is provided. The main focus is on the data flow of the FMM algorithm for which we develop the performance model.

\subsection{FMM Overview}
This overview is intended to introduce some key ingredients of the FMM. The mathematics behind the specific FMM kernels is outside the scope of this study, since it has very little to do with the communication model. For details of the mathematics we refer the reader to previous publications on FMM \cite{Beatson1997, Cheng1999}.

\subsubsection{Basic Component}
Both treecode \cite{Barnes1986} and FMM \cite{Greengard1987} are based on two key ideas: the tree representation for the spatial hierarchy, and the fast approximate evaluation.The spatial hierarchy means that the computational domain is hierarchically decomposed into increasing levels of refinement, and then the near and far subdomains can be identified at each level. The three-dimensional spatial domain of the treecode and FMM is represented by octrees, where the space is recursively subdivided into eight cells until the finest level of refinement or ``leaf level". Figure \ref{fig:tree} illustrates such a hierarchical space decomposition for a two-dimensional domain (a), associated to a quad-tree structure (b). The original FMM \cite{Greengard1988} is based on a series expansion of the Laplace Green's function$(1/r$ and therefore can be applied to the evaluation of related potentials and/or forces \cite{Gorn2004}. The approximation reduces the number of operations in exchange for accuracy.

\begin{figure*}[t]
\centering    
\includegraphics[width=0.8\textwidth]{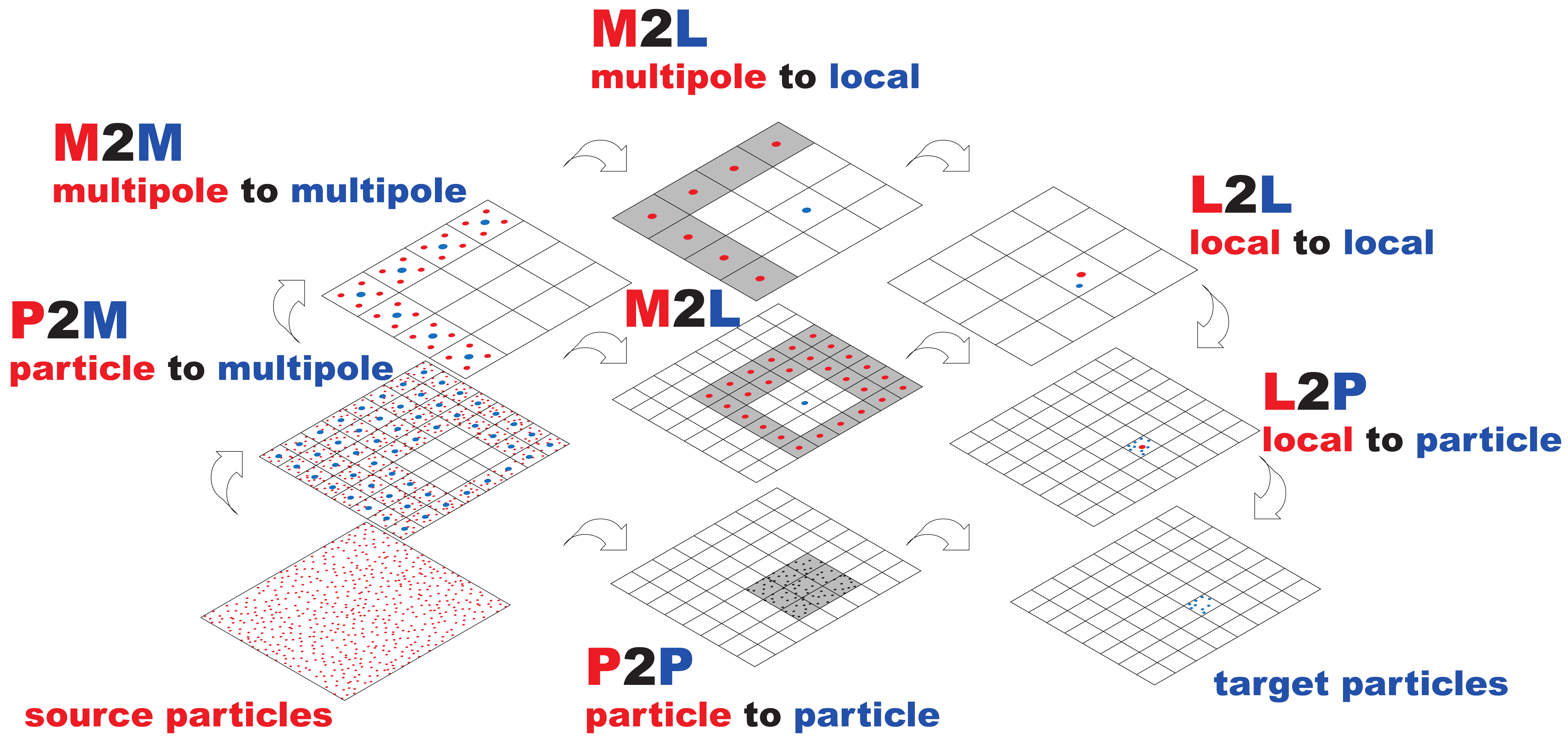}
\caption{Data-flow of FMM calculation. Data dependency is between red and blue points.}.
\label{fig:flow}
\end{figure*}

\subsubsection{Flow of Calculation}
Figure \ref{fig:flow}, shows the flow of FMM where the effect of the source particles, shown in red in the lower left corner, are calculated on the target particles, shown in blue in the lower right corner. The schematic is a 2-D representation of what is actually a 3-D octree structure. The calculation starts by transforming the mass/charge of the source particles to a multipole expansion (P2M). Then, the multipole expansion is translated to the center of larger cells (M2M). Then, the influence of multipoles on the particles is calculated in three steps. First, it translates the multipole expansion to a local expansion (M2L). Next, the center of expansion is translated to smaller cells (L2L). Finally, the effect of the local expansion in the far field is translated onto the target particles (L2P). All pairs interaction is used to calculate the effect of near field on target particles (P2P).

\subsection{FMM Communication Scheme}
\begin{figure*}[!t]
\centering
\includegraphics[width=0.95\textwidth]{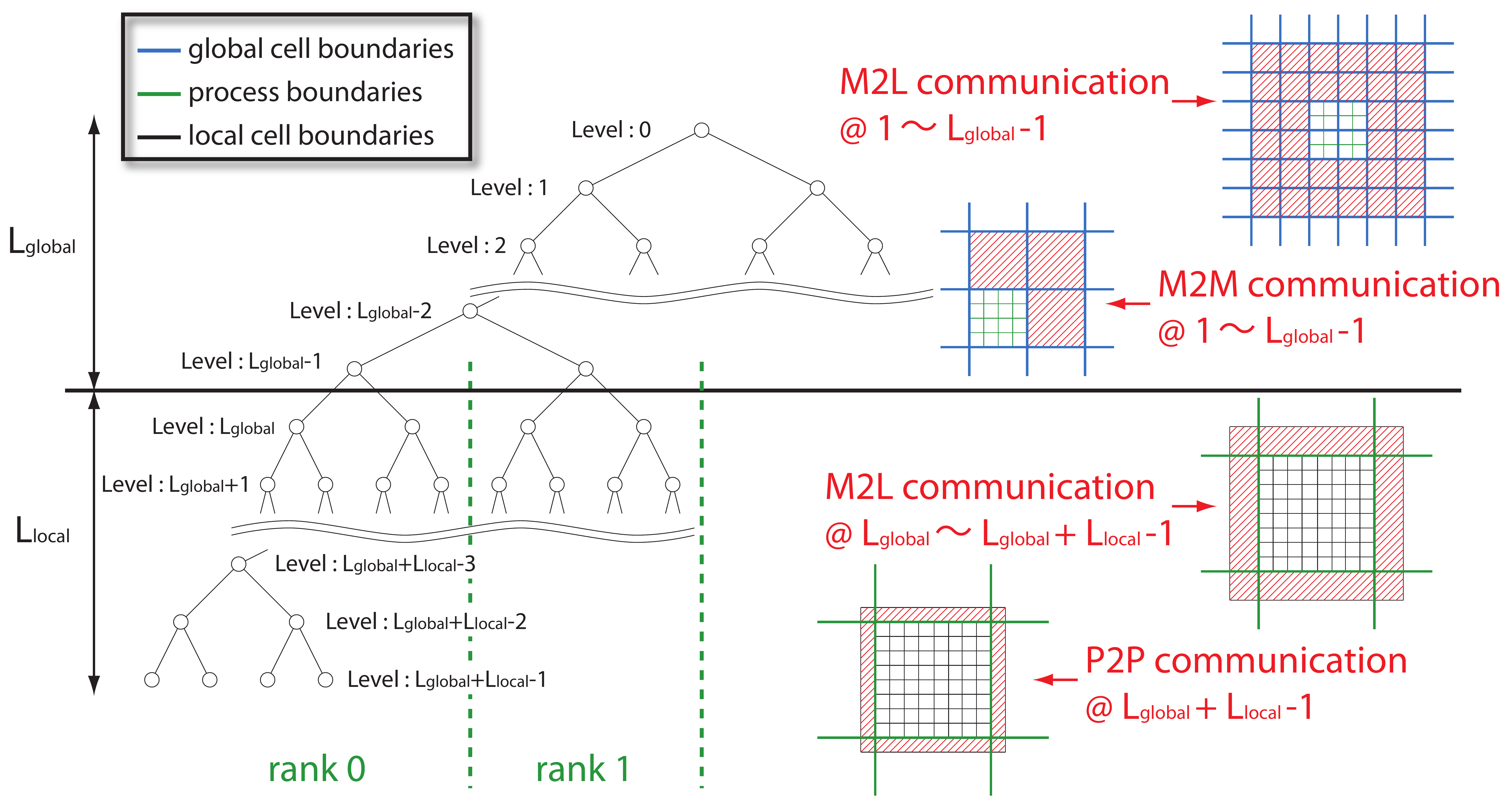}
\caption{Splitting of the local and global tree in FMM.}
\label{fig:global}
\end{figure*}

\begin{figure*}[!t]
\centering
\subfloat[Level=7]{\includegraphics[width=0.3\textwidth]{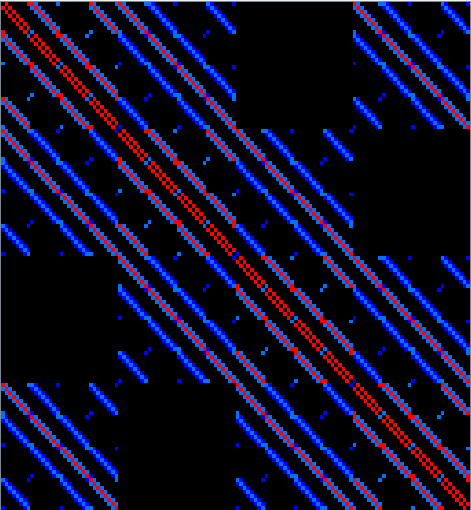}}
\hfil
\subfloat[Level=6]{\includegraphics[width=0.3\textwidth]{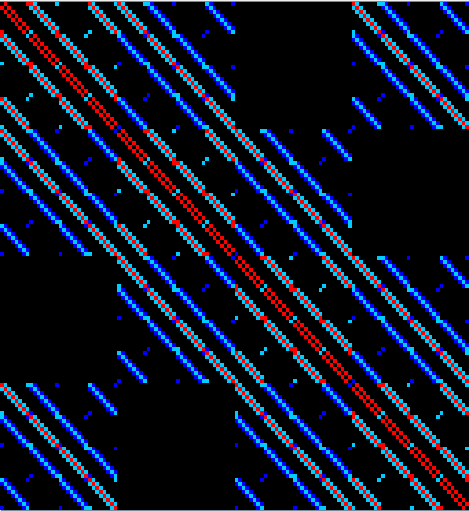}}
\hfil
\subfloat[Level=5]{\includegraphics[width=0.3\textwidth]{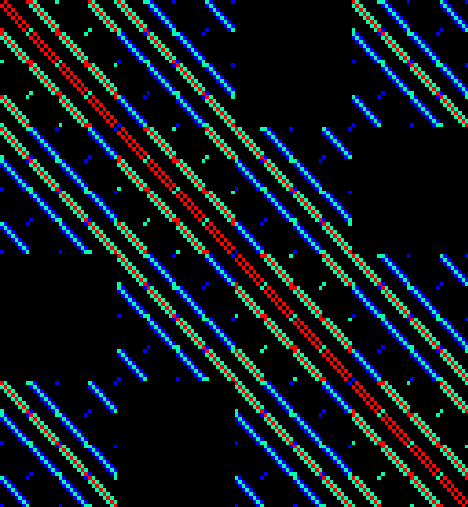}}
\hfil
\subfloat[Level=4]{\includegraphics[width=0.3\textwidth]{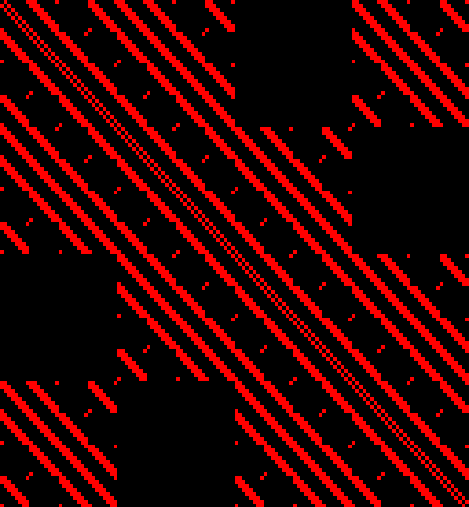}}
\hfil
\subfloat[Level=3]{\includegraphics[width=0.3\textwidth]{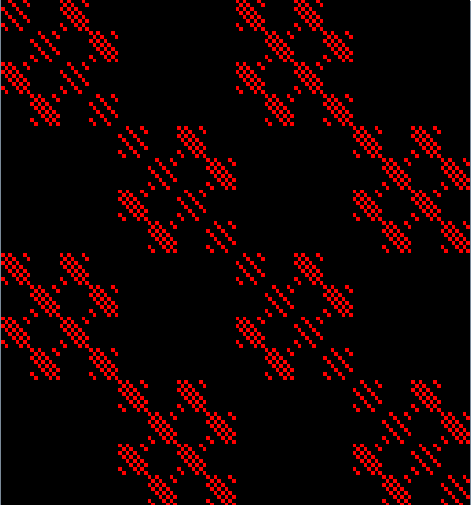}}
\hfil
\subfloat[Level=2]{\includegraphics[width=2in]{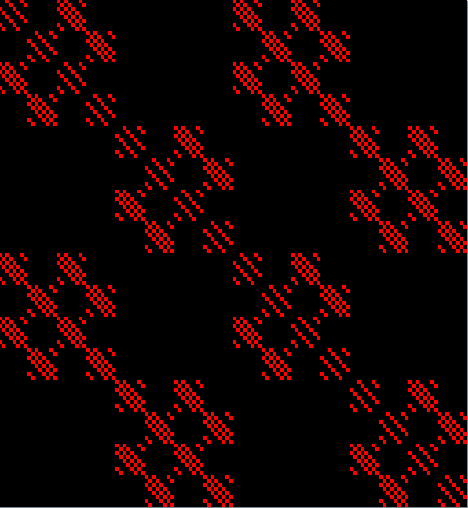}}
\caption{Level-by-level communication patterns for the M2L phase of an FMM with N=62,500 per process using 128 processes. Areas of black indicate zero messages between processes.}
\label{fig:pattern}
\end{figure*}

Partitioning of the FMM global tree structure and communication stencils is shown in Figure \ref{fig:global}. The binary tree on the left side is a simplification of what is actually an octree in a 3-D FMM. Likewise, the schematics on the right are a 2-D representation of what is actually a 3-D grid structure. Each leaf of the global tree is a root of a local tree in a particular MPI process, where the global tree has $L_{global}$ levels, and the local tree has $L_{local}$ levels. Each process stores only the local tree, and communicates the halo region at each level of the local and global tree as shown in the red hatched region in the four illustrations on the right. The blue, green, and black lines indicate global cell boundaries, process boundaries, local cell boundaries, respectively. The switch between local and global trees produces a change in the communication pattern as shown in Figure \ref{fig:pattern}.

\section{Modeling Performance}
Performance modeling is a key ingredient in high performance computing.  It has a great importance in the design, development and optimization of applications, architectures and communication systems. It also plays a crucial role in understanding important performance bottlenecks of complex systems. For this reason, performance models are used to analyze, predict, and calibrate performance for systems of interest. In this section we develop a performance model to understand the performance of FMM and to make predictions on future machines.

First, we start with the baseline model that is a combination of the latency and inverse bandwidth. We subsequently refine this baseline model to reach a complete model that is able to cover the relevant system architecture properties, with the exception that overlapping communication with computation is not considered in this work.

\begin{table}[b]
\caption{Amount of communication in FMM}
\centering
\begin{tabular}{lcc}
\hline
 & Cells to send / level & Total comm.\\
\hline
Global M2L & $26\times8$ & $\mathcal{O}(\log P)$\\
Global M2M & 7 & $\mathcal{O}(\log P)$\\
Local M2L & $(2^i+4)^3-8^i$ & $\mathcal{O}((N/P)^{2/3})$\\
Local P2P & $(2^i+2)^3-8^i$ & $\mathcal{O}((N/P)^{2/3})$\\
\hline
\end{tabular}
\label{tab:fmmcomm}
\end{table}

\subsection{FMM Communication Phases}\label{sec:phases}
As shown in Figure \ref{fig:global}, our FMM uses a separate tree structure for the local and global tree. In order to construct a performance model for the communication in FMM, we estimate the amount of data that must be sent at each level of the hierarchy. Table \ref{tab:fmmcomm} shows the number of cells that are sent, which correspond to the illustrations in Figure 3. $L_{global}$ is the depth of the global tree, $L_{local}$ is the depth of the local tree. We define $N$ as the global number of particles, and $P$ as the number of processes (MPI ranks). The global tree is constructed so that each MPI process is a leaf node in the global tree. Therefore, the depth of the global tree only depends on the number of processes $P$ and not $N$. The depth of the global tree grows with $\log_8P$, whereas the depth of the local tree grows with $\log_8(N/P)$. For the current calculations we are assuming a nearly uniform particle distribution (as in explicit solvent molecular dynamics) and therefore a full octree structure.

\subsubsection{Global M2L}
In Table \ref{tab:fmmcomm} we show the amount of cells to send per level and the total amount of communication for all levels. There are four types of communication in our FMM, which correspond to the four stages shown with the red hatching in Figure \ref{fig:global}. The first is the ``Global M2L" communication, which sends $26\times8$ cells at each level, as shown at the top right of Figure \ref{fig:global}. The green lines are the process boundaries and the blue lines are the cell boundaries, which means one FMM cell belongs to many processes in the global tree. In order to avoid redundant communication, we index each process that shares a global cell and perform a one-to-one communication between the processes with matching indices only. In order to further reduce the communication, we select one process for a group of eight cells to do the communication. Therefore, the number of processes to communicate with ($p_i$) is always $26$ and the number of cells to send is always $8$ for every process and for every level in the global tree. In other words, for the ``Global M2L" communication the message size and number of sends is constant regardless of $N$ and $P$, and only the number of hops between the processes will increase depending on the network topology. On torus networks, we map the MPI ranks to the torus and synchronize the direction of the $26$ one-to-one communications. The communication per level is $\mathcal{O}(1)$ and the number of levels in the global tree is $\mathcal{O}(\log P)$, so the total communication complexity for this stage is $\mathcal{O}(\log P)$ as shown in Table \ref{tab:fmmcomm}.

\subsubsection{Global M2M}
The second type of communication is the ``Global M2M", which sends $7$ cells at each level, as shown in Figure \ref{fig:global}. We use a similar technique to the ``Global M2L" case to avoid redundant communication by pairing the MPI ranks for the one-to-one communication when many processes share the same global cell. The number of processes to communicate with is always seven and the number of cells to send is always one for every process and for every level in the global tree. Similar to the ``Global M2L" case, only the number of hops during the one-to-one communication will increase, and the rate depends on the network topology. The communication per level is $\mathcal{O}(1)$ and the number of levels is $\mathcal{O}(\log P)$, so the total communication is $\mathcal{O}(\log P)$ for the ``Global M2M" stage.

\subsubsection{Local M2L}
The third type of communication is the ``Local M2L", which is shown in the red hatching in the second picture from the bottom on the right side of Figure \ref{fig:global}. The process boundaries shown in green are coarser than the local cell boundaries shown in black, which means that one process contains many cells contrary to the previous two communication types. In a full octree structure, we know that all cells are non-empty so we simply need to send two layers of halo cells for the M2L calculation at each level, as shown in Figure \ref{fig:global}. Therefore, the number of processes to communicate with is always the $26$ neighbors, and the number of cells to send depends on the level. At level $i$ of the local tree, there are $2^i$ cells in each direction. Two layers of halo cells on each size will create a volume of $(2^i+4)^3$ cells, and subtracting the center volume $8^i$ will give $(2^i+4)^3-8^i$ as shown in Table \ref{tab:fmmcomm}. The leading term is $\mathcal{O}(4^i)$ since the $8^i$ term cancels out. Since the number of levels in the local tree grow as $\log_8(N/P)$ the communication complexity for the ``Local M2L" is $\mathcal{O}(4^{\log_8(N/P)})=\mathcal{O}((N/P)^{2/3})$. This can also be understood as the surface to volume ratio of the bottom two illustrations in Figure \ref{fig:global}. Since $N/P$ is constant for weak scaling and decreases for strong scaling, this part does not affect the asymptotic weak/strong scalability of the FMM.

\subsubsection{Local P2P}
The fourth type of communication in the FMM is the ``Local P2P", which is town in the bottom picture on the right side of Figure \ref{fig:global}. This communication only happens at the bottom level of the local tree. Similar analysis to the ``Local M2L" stage shows that $(2^i+2)^3-8^i$ cells must be sent, as shown in Table \ref{tab:fmmcomm}. In this case, $i$ is exactly $\log_8(N/P)$ and we obtain the same asymptotic amount of communication of $\mathcal{O}((N/P)^{2/3})$. Similar to the ``Local M2L", this part does not affect the asymptotic weak/strong scalability of the FMM. However, the content of the data is different from the previous three cases where the multipole expansion coefficients were being sent. In the P2P communication the coordinates and the charges of every particle that belongs to the cell must be sent. Therefore, the asymptotic constant of $\mathcal{O}(N/P)^{2/3}$ is typically much larger than that of the ``Local M2L", and this could be the dominant part of the communication time depending on the number of particles per leaf cell.

\subsection{Baseline Model ($\alpha-\beta$ model)}
To model interprocess communication, we start by the basic $\alpha-\beta$ model, where $\alpha$ represents communication latency, where $\beta$ is the send time per-Byte (inverse bandwidth). Using the basic $\alpha-\beta$ model, a message send cost can be represented as
\begin{equation}
T_{\alpha-\beta}=\alpha+n\beta
\end{equation} 
where $n$ is the number of Bytes in the message.

This basic model describes the communication over an ideal architecture where the communication cost does not depend on processor locations or network traffic caused by many processors communicating at the same time \cite{Foster1995}. For a more realistic architecture, a more detailed model is needed. For this reason, we add penalties to this basic model to take into account machine-specific performance issues. In particular, we consider communication distance, interconnection switching delay, limited bandwidth, and the effect of multiple cores on a single node contending for available resources.

\subsection{Distance Penalty ($\alpha-\beta-\gamma$ Model)}
Following \cite{Gahvari2011}, we refine the assumption that distance between processors in interconnected networks does not have effect on communication time. To take into account the effect of distance we refine the baseline model according to the number of extra hops a message travels
\begin{equation}
T_{\alpha-\beta-\gamma}=\alpha+n\beta+(h-h_{m})\gamma,
\end{equation}
where $h$ is the number of hops a message travels, $h_m$ is the smallest possible number of hops a message can travel in the network, and $\gamma$ is the delay per extra hop. If there is no network contention and all messages travel with minimum number of hops, this distance penalty should have no effect.

\subsection{Bandwidth Penalty on $\beta$}
The peak hardware bandwidth is rarely achieved in message passing. Therefore, we multiply $\beta$ by $B_{max}/B$ to incorporate the ratio between the peak hardware per-node bandwidth $B_{max}$ and the effective bandwidth from the benchmark $B$.
\begin{equation}
T_{\beta-Penalty}=\alpha+n\beta\frac{B_{max}}{B}+(h-h_{m})\gamma
\end{equation}

\subsection{Multicore Penalty on $\alpha$ or $\gamma$}
Increasing the number of cores per node increases the data traffic between nodes, and could potentially result in congestion. Furthermore, larger number of cores per node introduces more noise caused by access to resources shared by multiple cores. To model these effects, we multiply $\alpha$ and/or $\gamma$ by the number of active cores per node $c$. This model focuses on the worst case behavior where a machine's aggregate bandwidth could be exceeded by all cores communicating simultaneously. The resulting models are
\begin{eqnarray}
T_{\alpha-Penalty}=c\alpha+n\beta+(h-h_{m})\gamma\\
T_{\gamma-Penalty}=\alpha+n\beta+c(h-h_{m})\gamma
\end{eqnarray}

\section{Model Validation}

\subsection{Machine Description}
To validate our performance models, we benchmark our FMM code on three different architectures; Shaheen, Mira, and Titan.

\textbf{Shaheen} is $16$ racks of an IBM BlueGene/P. Each rack contains $1024$ PowerPC 450 CPUs with $4$ cores running at $850$MHz with $32$kB private L1 cache and $8$MB shared L3 cache. Each compute node has $2$GB RAM with $13.6$ GB/s memory bandwidth. The nodes are connected by 3-D torus network with $5.1$GB/s injection bandwidth per node.

\textbf{Mira} is $48$ racks of an IBM BlueGene/Q. Each rack contains $1024$ Power A2 CPUs with $16+1$ cores running at $1.6$GHz with $16$kB private L1 cache and $32$MB shared L2 cache. Each compute node has $16$GB RAM with $42.6$GB/s memory bandwidth. The nodes are connected by a 5-D torus network with $20$GB/s injection bandwidth per node.

\textbf{Titan} is a Cray XK7 system with $18,688$ compute nodes each equipped with an AMD Opteron 6274 CPU and NVIDIA Kepler K20X GPU. The CPU has $16$ cores running at $2.2$ GHz with $16$ kB L1 cache, $2\times4$ MB L2 cache, and $8\times2$ MB L3 cache. The GPU has $15\times64$ cores running at $730$ MHz  with $64+48$ kB L1 cache and $1.5$ MB L2 cache. Each compute node has $32$ GB of RAM with $51.2$ memory bandwidth. The nodes are connected by a 3-D torus with $20$GB/s of injection bandwidth per node. We do not use any of the GPUs in the current study.

In order to obtain the machine parameters, the \texttt{b\_eff} benchmark in the HPC Challenge suite \cite{Luszczek2005} was used to determine the parameters $\alpha$ and $\beta$. We report the best-case latency and bandwidth measurements. To find the parameter $\gamma$, we followed the same procedure as Gahvari \textit{et al.} \cite{Gahvari2011}. The machine parameters for Shaheen, Mira, and Titan are shown in Table \ref{tab:parameters}. Note that our definition of $\beta$ is defined as send time per Byte, whereas Gahvari \textit{et al.} define their $\beta$ as send time per element (8 Bytes).

\begin{table}[!t]
\caption{Machine parameters for latency $\alpha$, inverse bandwidth $\beta$, and distance penalty $\gamma$, on Shaheen, Mira, and Titan.}
\label{tab:parameters}
\centering
\begin{tabular}{|c||c|c|c|c|}
\hline
& Shaheen & Mira & Titan\\
\hline
$\alpha$ & $4.12\ \mu s$ & $5.33\ \mu s$ & $1.67\ \mu s$\\
\hline
$\beta$ & $2.14\ ns$ & $1.32\ ns$ & $1.62\ ns$\\
\hline
$\gamma$ & $29.9\ ns$ & $134\ ns$ & $284\ ns$\\
\hline
\end{tabular}
\end{table}

\subsection{Experimental Setup}
We ran the FMM code for 10 steps and measured the time spent on the communication for the ``Global M2L" and ``Local M2L" phases. The results are then divided by 10 to get the average time spent at each level. The ``Global M2M" phase was negligible and the ``Local P2P" phase only occurs at the bottom level and is irrelevant to the scalability of the FMM, so we do not consider these two phases in the current analysis. We used the Laplace kernel in three dimensions with random distribution of particles in a cube. We use periodic boundary conditions so that there is no load imbalance at the edges of the domain. The number of MPI processes was varied between $P=\{128, 1024, 8192\}$, while the number of particles per process was kept constant at $N/P=62,500$.

\begin{table}[!t]
\caption{Statistics of the M2L communication.}
\label{tab:statistics}
\centering
\begin{tabular}{|c|c|c|c|c|c|}
\multicolumn{4}{c}{(a) 128 Processes}\\
\hline
Level & Cells & Sends & Bytes\\
\hline
0 & 1 & 0 & 0\\
\hline
1 & 8 & 0 & 0\\
\hline
2 & 64 & 26 & 46592\\
\hline
3 & 512 & 26 & 46592\\
\hline
4 & 4096 & 26 & 46592\\
\hline
5 & 32768 & 26 & 100352\\
\hline
6 & 262144 & 26 & 272384\\
\hline
7 & 2097152 & 26 & 874496\\
\hline
\end{tabular}\\
\vspace{2mm}

\begin{tabular}{|c|c|c|c|c|c|}
\multicolumn{4}{c}{(b) 1024 Processes}\\
\hline
Level & Cells & Sends & Bytes\\
\hline
0 & 1 & 0 & 0\\
\hline
1 & 8 & 0 & 0\\
\hline
2 & 64 & 26 & 46592\\
\hline
3 & 512 & 26 & 46592\\
\hline
4 & 4096 & 26 & 46592\\
\hline
5 & 32768 & 26 & 46592\\
\hline
6 & 262144 & 26 & 100352\\
\hline
7 & 2097152 & 26 & 272384\\
\hline
8 & 16777216 & 26 & 874496\\
\hline
\end{tabular}\\
\vspace{2mm}

\begin{tabular}{|c|c|c|c|c|c|}
\multicolumn{4}{c}{(c) 8192 Processes}\\
\hline
Level & Cells & Sends & Bytes\\
\hline
0 & 1 & 0 & 0\\
\hline
1 & 8 & 0 & 0\\
\hline
2 & 64 & 26 & 46592\\
\hline
3 & 512 & 26 & 46592\\
\hline
4 & 4096 & 26 & 46592\\
\hline
5 & 32768 & 26 & 46592\\
\hline
6 & 262144 & 26 & 46592\\
\hline
7 & 2097152 & 26 & 100352\\
\hline
8 & 16777216 & 26 & 272384\\
\hline
9 & 134217728 & 26 & 874496\\
\hline
\end{tabular}
\end{table}

Table \ref{tab:statistics} shows communication information and statistics when running the FMM on $128$, $1024$, and $8192$ processes. ``Level" is the level within the tree structure and goes from $0$ to $L_{global}+L_{local}-1$, where $L_{local}=4$ for $N/P=62,500$. Therefore, the bottom four levels in Table \ref{tab:statistics} (a), (b), and (c) belong to the local tree. The depth of the global tree $L_{global}$ is $4$, $5$, and $6$ for $128$, $1024$, and $8192$ processes, respectively. ``Cells" is the total number of cells at that level of the tree structure, which is simply $8^{Level}$ for a full octree. ``Sends" is the number of processes that each processes sends to. As mentioned in Section \ref{sec:phases} we have developed a communication scheme that limits the number of sends to $26$ regardless of the problem size, number of processes, or the level. ``Bytes" is the aggregate data size that is sent by a given process at each level of the tree. As shown in Table \ref{tab:fmmcomm}, the number of cells for the ``Global M2L" communication is $26\times8$. For each cell we are sending $56$ multipole expansion coefficients in single precision ($4$ Bytes). Therefore, the total number of Bytes for the ``Global M2L" phase is $26\times8\times56\times4=46592$. We can see from Table \ref{tab:fmmcomm} that the amount of cells involved in the ``Local M2L" communication can be calculated by $(2^i+4)^3-8^i$, where $i$ is the level in the local tree (not the ``Level" shown in Table \ref{tab:statistics}). For example, for level one in the local tree, the amount of cells will be $(2^1+4)^3-8^1$ which is equivalent to $26\times8$. This is why the ``Bytes" is the same for the ``Global M2L" and the first level of the ``Local M2L" in Table \ref{tab:statistics}.

\subsection{Model Validation}
We compare the actual communication time for the M2L communication with our performance model on Shaheen, Mira, and Titan. We compare against same combination of models as in the multigrid study \cite{Gahvari2011}. The combinations are:

\begin{enumerate}
\item Baseline model ($\alpha-\beta$ model)
\item With distance penalty ($\alpha-\beta-\gamma$ model)
\item With distance and bandwidth penalty ($\beta$ penalty)
\item With distance and bandwidth penalty, plus multicore penalty on latency ($\alpha,\beta$ penalty)
\item With distance and bandwidth penalty, plus multicore penalty on distance ($\beta,\gamma$ penalty)
\item With distance and bandwidth penalty, plus multicore penalty on latency and distance ($\alpha,\beta,\gamma$ penalty)
\end{enumerate}

\begin{figure}
\centering
\subfloat[128 processes]{\includegraphics[width=0.5\textwidth]{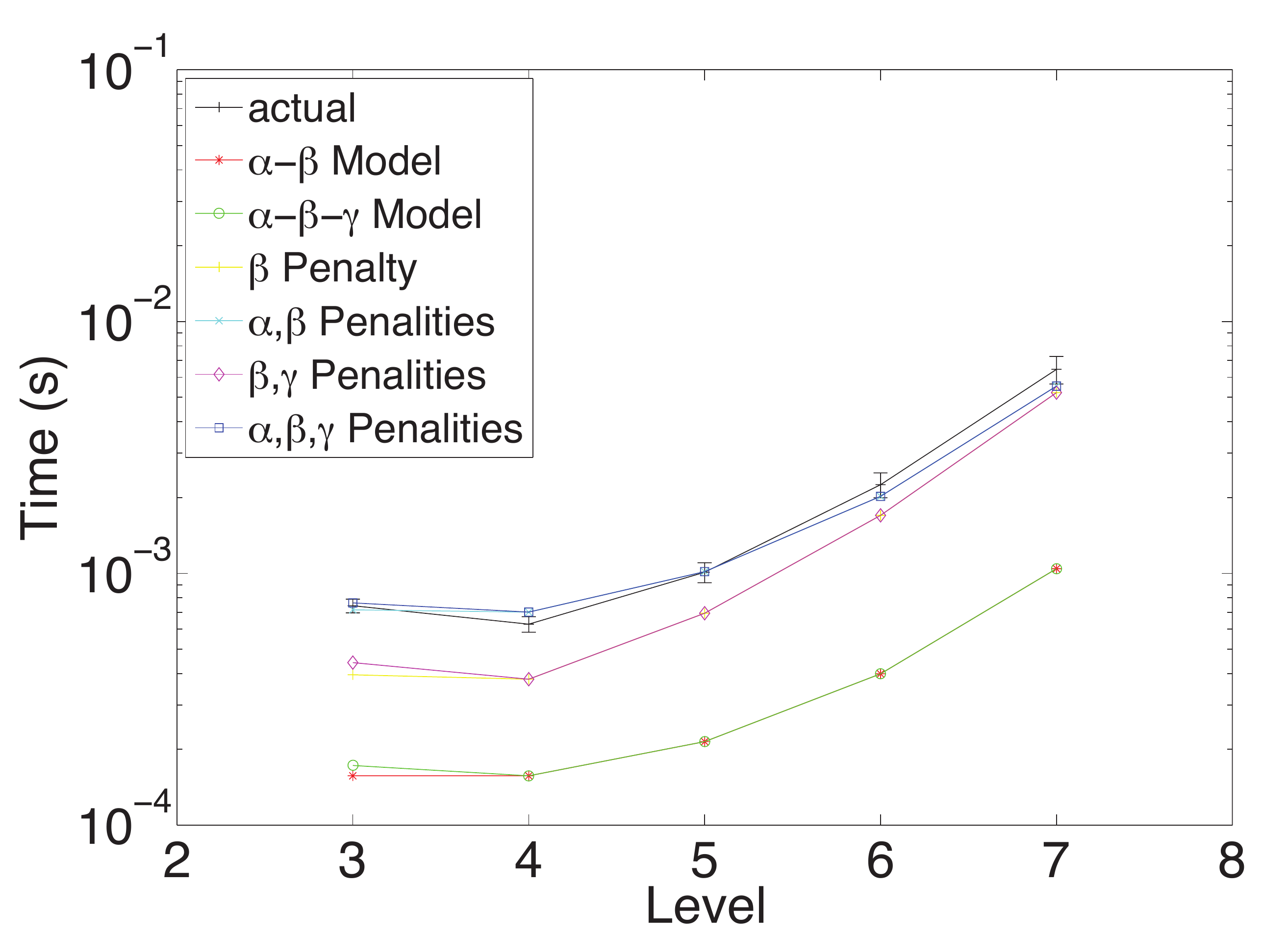}}\\ 
\subfloat[1024 processes]{\includegraphics[width=0.5\textwidth]{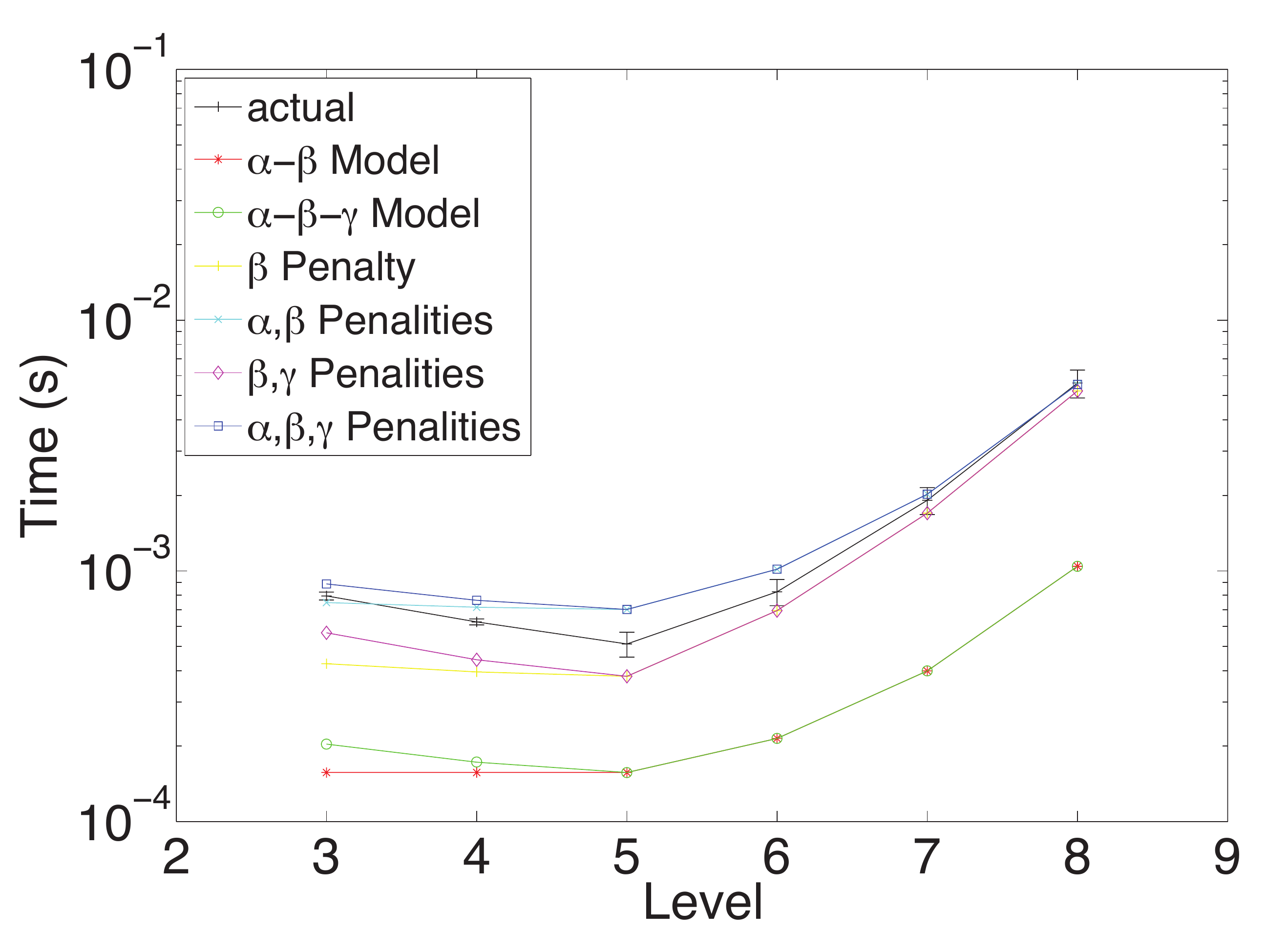}}\\
\subfloat[8192 processes]{\includegraphics[width=0.5\textwidth]{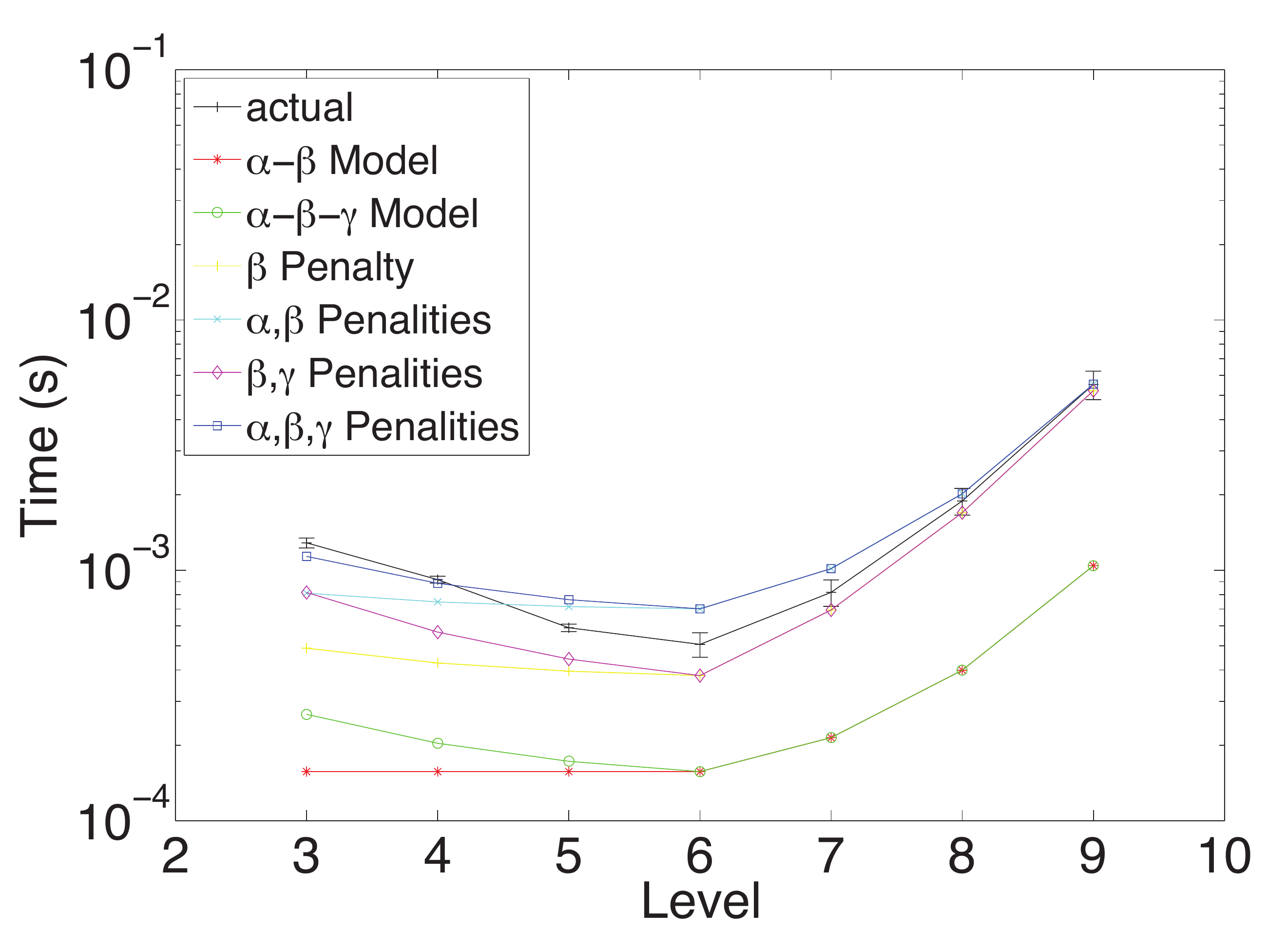}}
\caption{Performance model prediction and actual time for M2L communication phase on Shaheen.}
\label{fig:shaheen}
\end{figure}

The results on Shaheen are shown in Figure \ref{fig:shaheen}. The actual measured performance is shown as a black line, where an error bar is drawn according to the standard deviation in communication time among the different MPI ranks. By comparing the Bytes in  Table \ref{tab:statistics} with the communication time in Figure \ref{fig:shaheen}, we see that the deepest four levels that belong to the ``Local M2L" phase have a communication time that is proportional to the data size being sent. The main discrepancy in the models is caused by the $\beta$ penalty, for which the ratio between the theoretical injection bandwidth and the \texttt{b\_eff} benchmark results is accounted for. The actual communication time agrees well with the models with $\alpha$, $\beta$, and $\gamma$ penalties. For the shallow levels that belong to the ``Global M2L" phase, the communication time increases as the level decreases/coarsens. The reason for this can be understood by looking back at Figure \ref{fig:global}, where the ``Global M2L" is communicating with farther processes at coarser levels of the tree. Since we are mapping the geometric partitioning of the octree to the 3-D torus network of Shaheen, the proximity in the octree directly translates to the proximity in the network. Therefore, even though the data size is constant for all levels in the ``Global M2L" phase, the number of hops is larger, which accounts for switching delays and also network contention to some extent. This increases the communication time at coarser levels and the models that incorporate $\gamma$ are able to predict this behavior.

\begin{figure}
\centering
\subfloat[128 processes]{\includegraphics[width=0.5\textwidth]{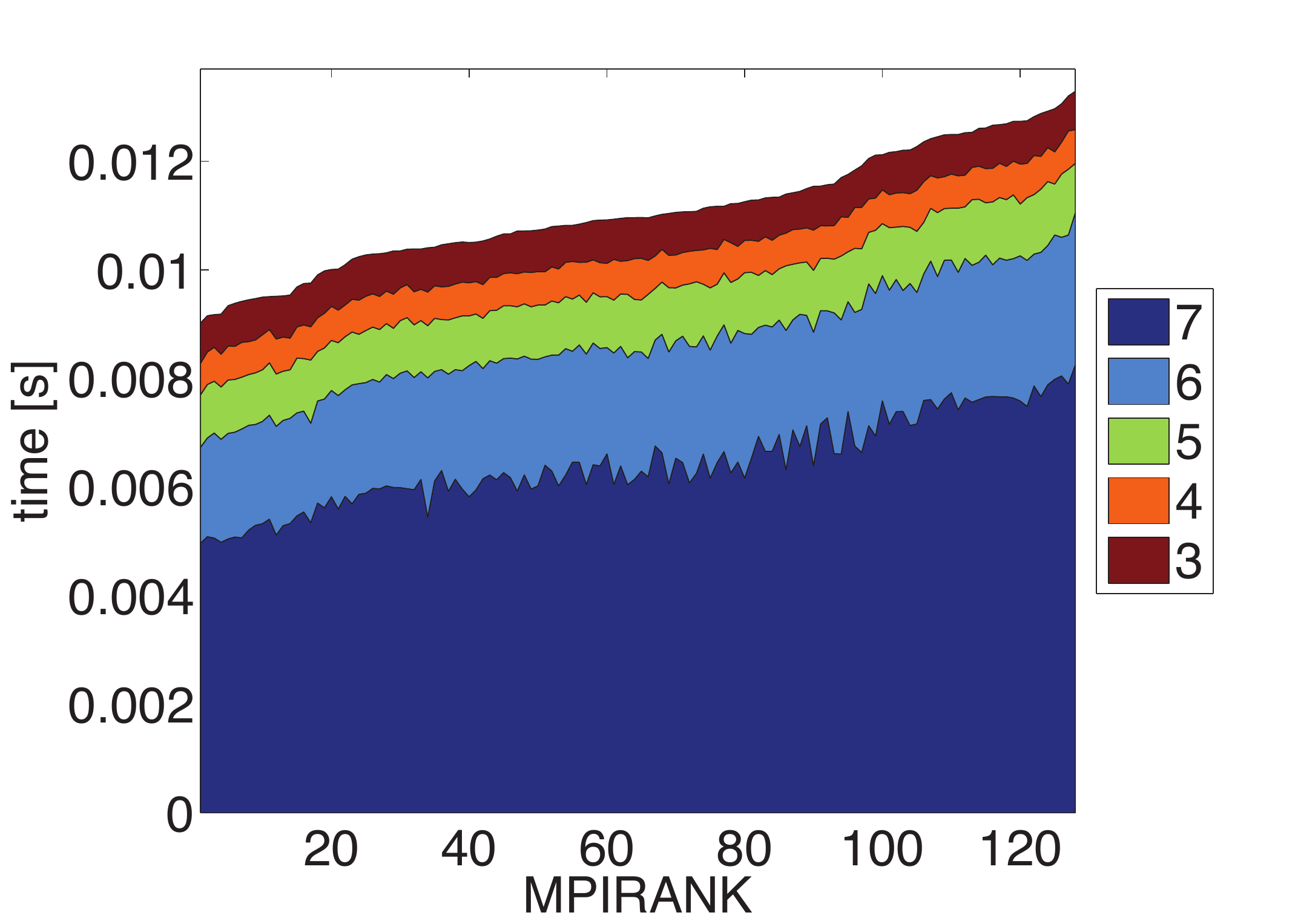}}\\ 
\subfloat[1024 processes]{\includegraphics[width=0.5\textwidth]{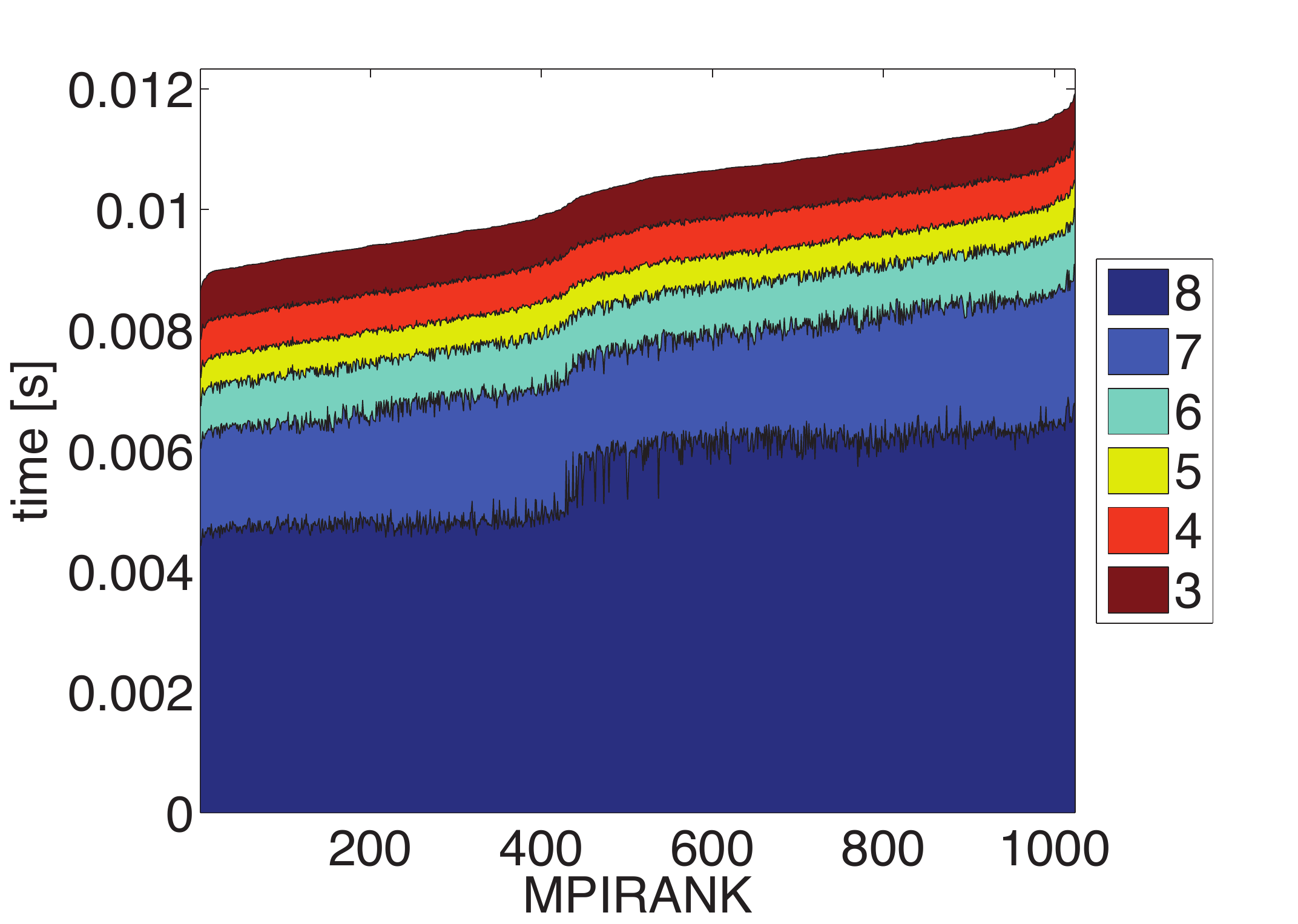}}\\
\subfloat[8192 processes]{\includegraphics[width=0.5\textwidth]{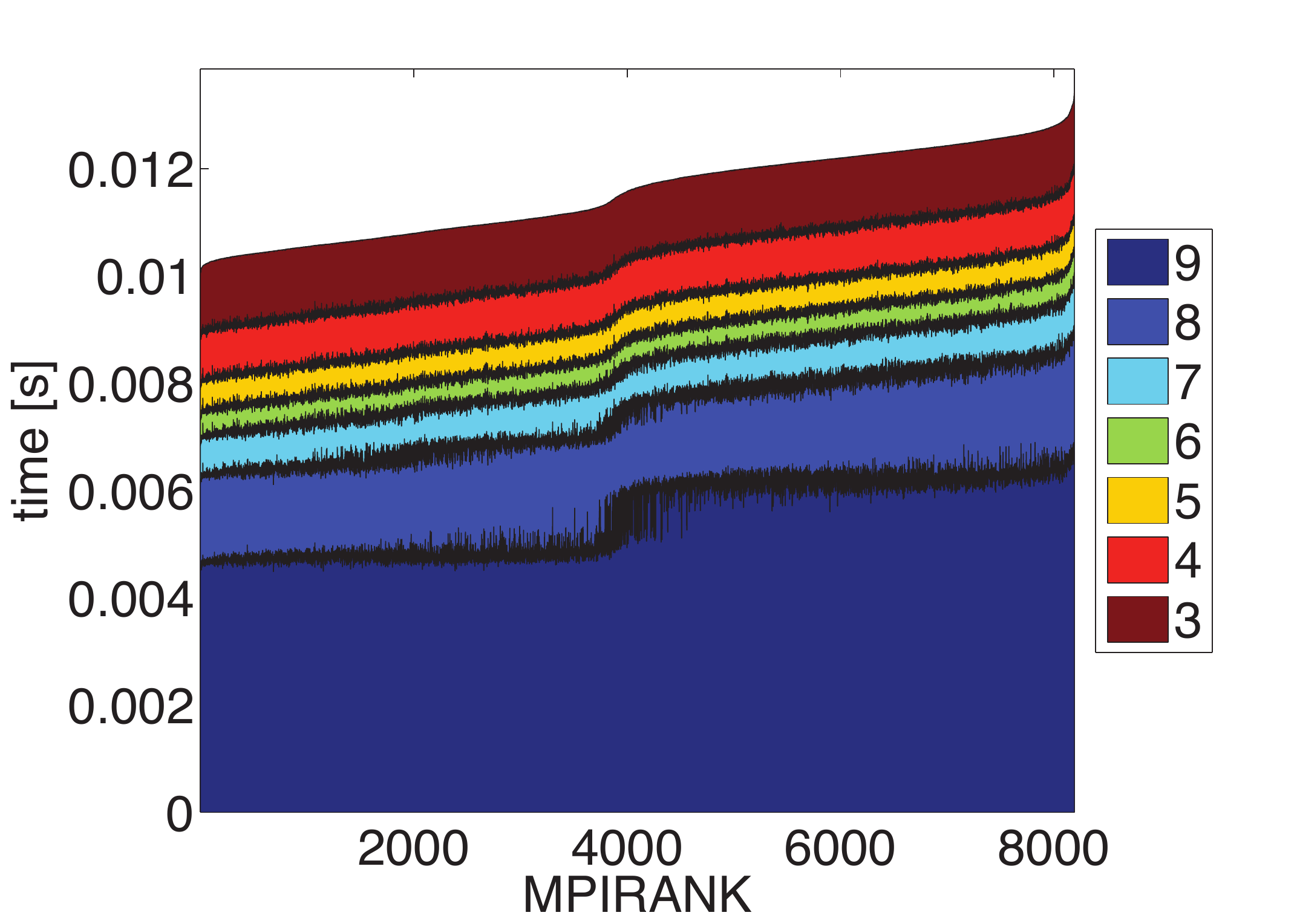}}
\caption{Load balance of M2L communication phase on Shaheen.}
\label{fig:load}
\end{figure}

In Figure \ref{fig:load}, the M2L communication time on Shaheen is plotted against the MPI rank to show the load balance between the processes. Each color shows M2L communication at a different level of the tree structure, and the numbers in the legend represent the levels. The communication time of each level is stacked on top of each other so that the total hight of the area plot represents the total M2L communication time shown in Figure \ref{fig:shaheen}. The MPI ranks are sorted according to the total M2L communication time for better visibility in the small differences between processes. As can be seen from the figure, the load balance is quite good. The imbalance seems to come from the finest levels, which are 7, 8, and 9 for 128, 1024, and 8192 processes, respectively.

\begin{figure}
\centering
\subfloat[128 processes]{\includegraphics[width=0.45\textwidth]{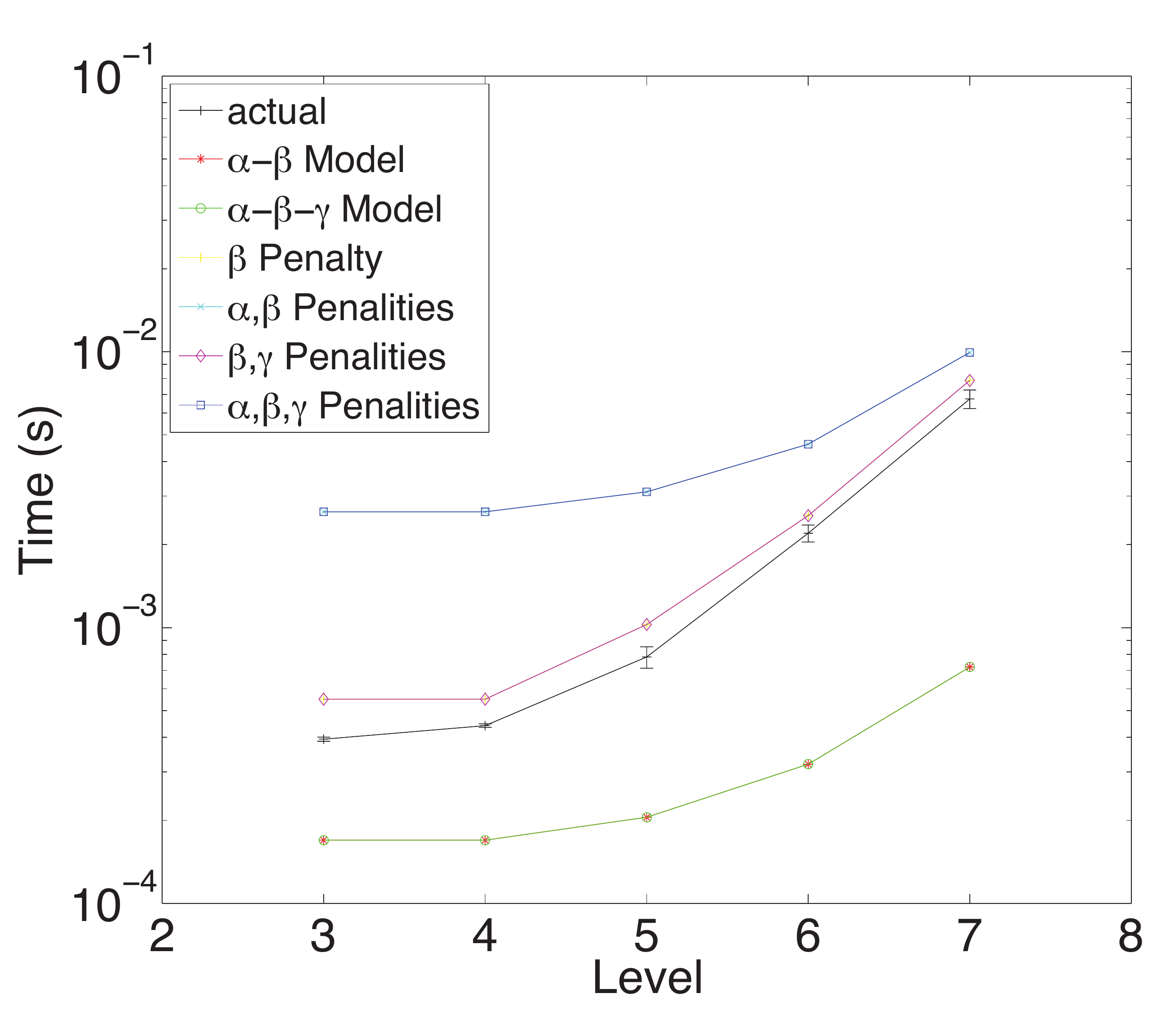}}\\ 
\subfloat[1024 processes]{\includegraphics[width=0.45\textwidth]{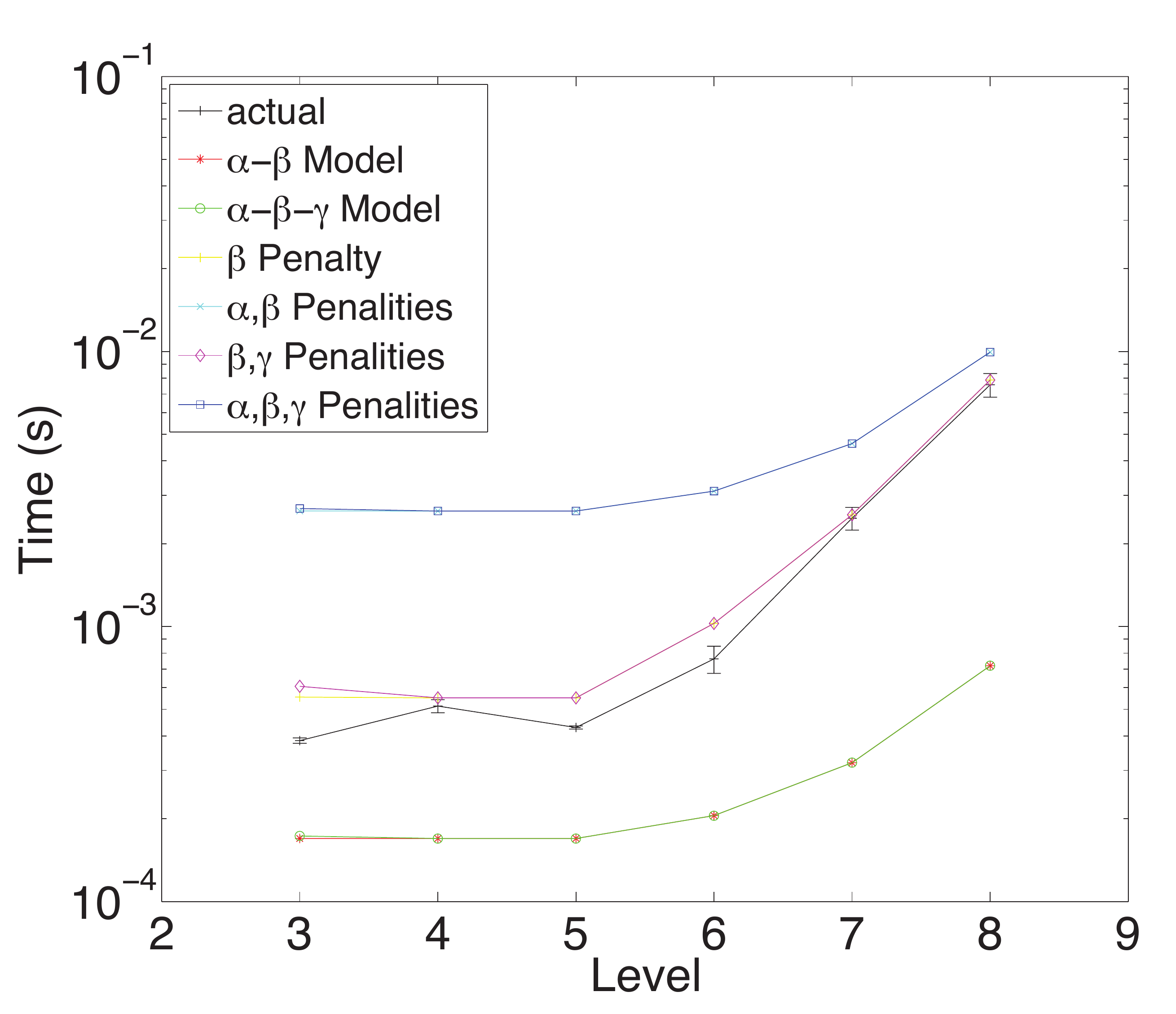}}\\
\subfloat[8192 processes]{\includegraphics[width=0.45\textwidth]{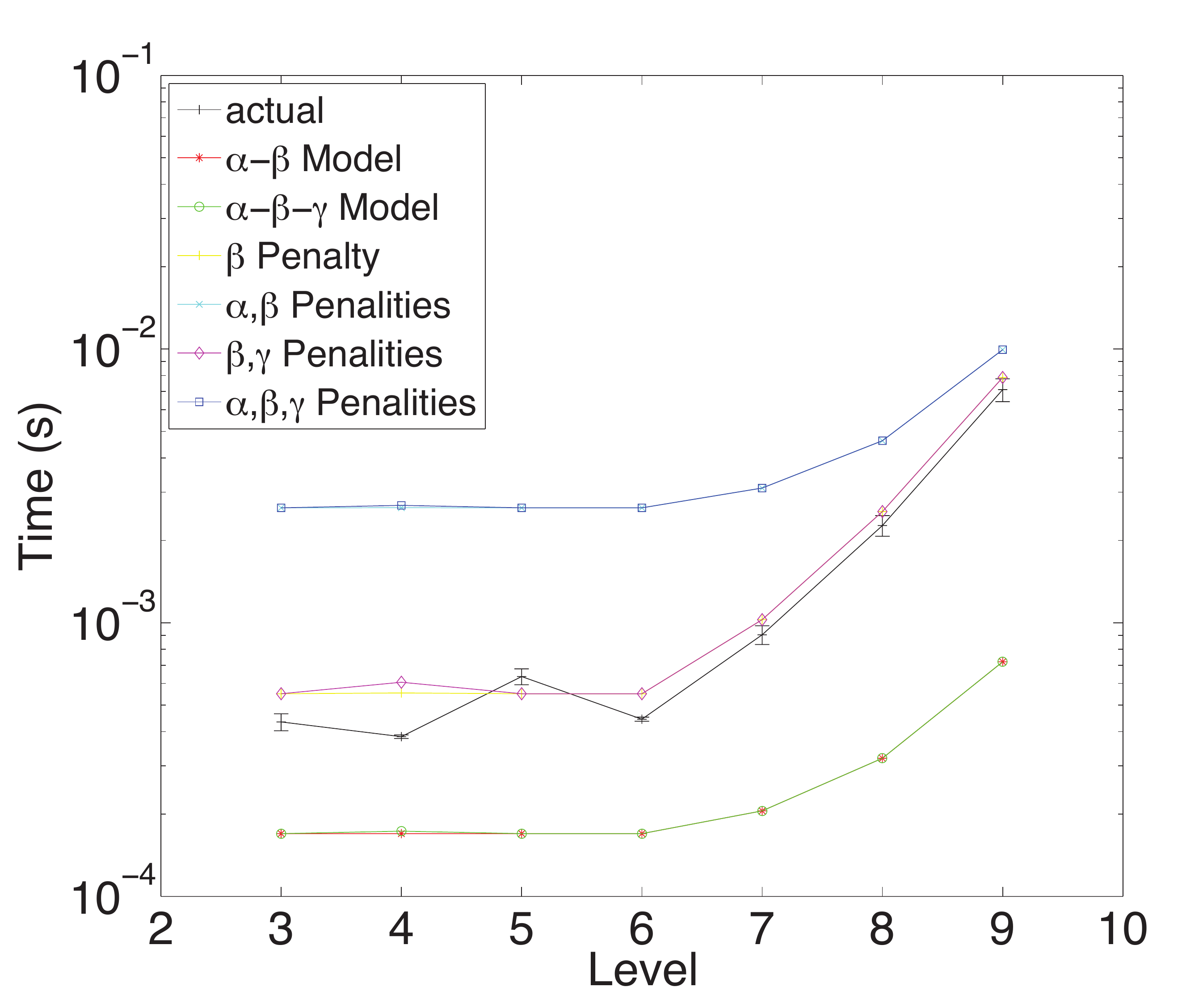}}
\caption{Performance model prediction and actual time for M2L communication phase on Mira.}
\label{fig:mira}
\end{figure}

The M2L communication time on Mira is plotted along with the six model predictions in Figure \ref{fig:mira}. Similar to the runs on Shaheen, the main difference in the model predictions is caused by the $\beta$ penalty. We also see a discrepancy between the model predictions with and without the $\alpha$ penalty for the ``Global M2L" phase (coarser levels). The multicore penalty is very small on the Bluegene/Q, so the model without the multicore penalty show a better agreement in terms of slope of the curve. For example ``$\beta$ Penalty" and ``$\beta$ Penalty" match quite well with the actual time. This lack of multicore penalty has been observed in other applications where the use of hybrid OpenMP+MPI approach did not improve the performance over a flat MPI approach \cite{lee2013}. Contrary to the runs on Shaheen, the communication time has a nearly flat profile for the ``Global M2L" phase. This is because the 5-D torus network minimizes the number of hops and network contention so the degradation at coarse levels of the tree is minimal. Far nodes in the octree are not so far in the Bluegene/Q network topology.

\begin{figure}
\centering
\subfloat[128 processes]{\includegraphics[width=0.45\textwidth]{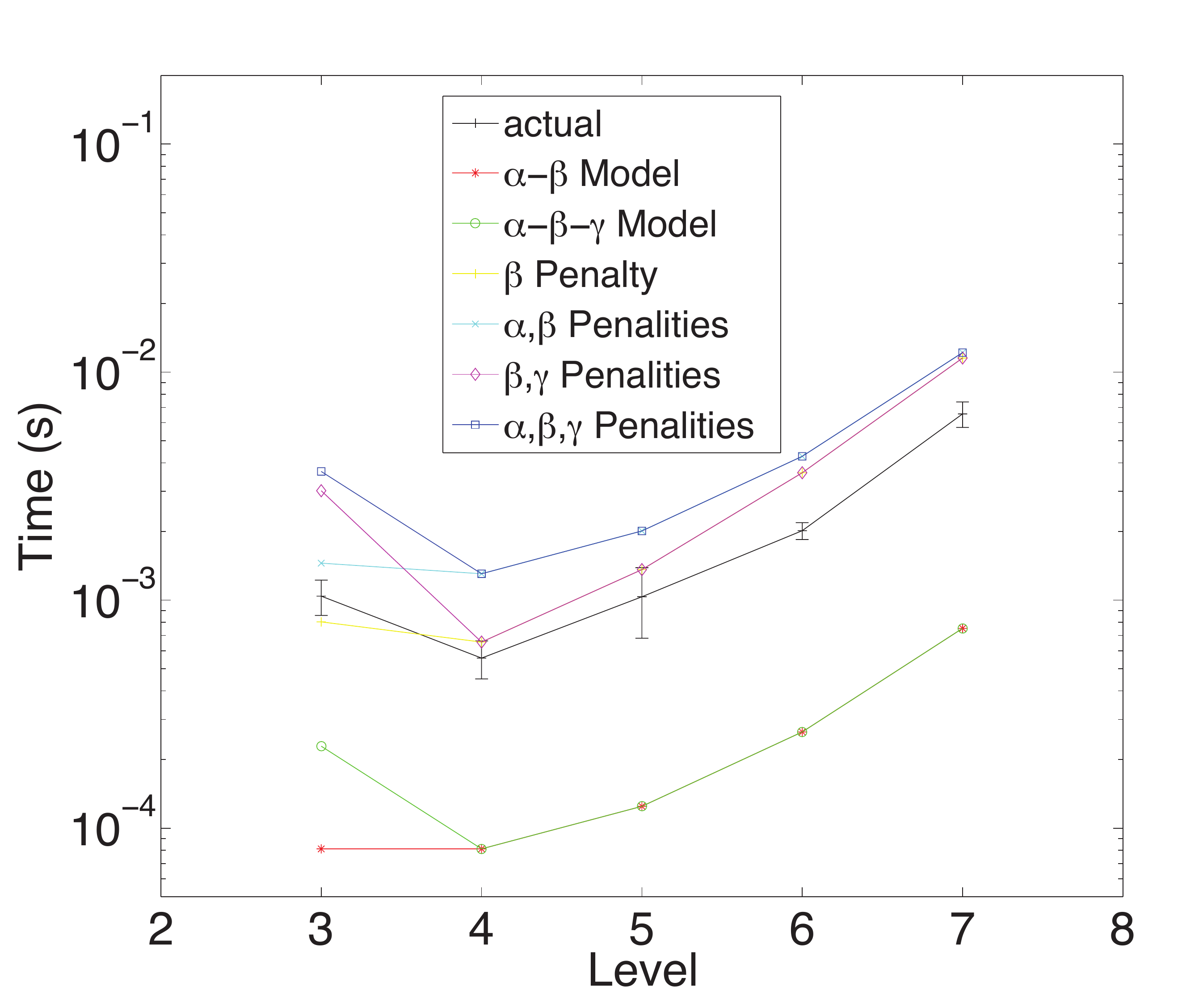}}\\ 
\subfloat[1024 processes]{\includegraphics[width=0.45\textwidth]{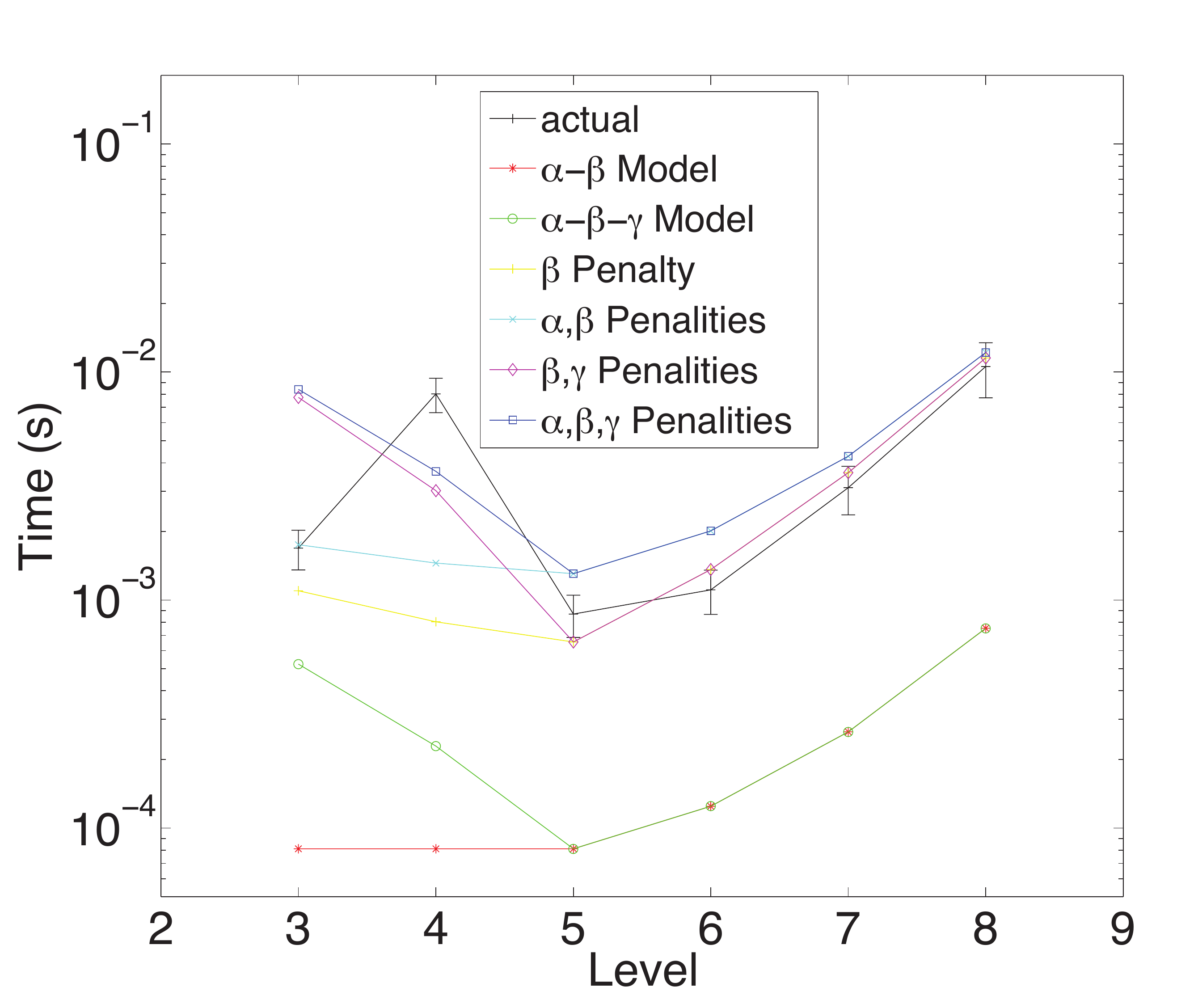}}\\
\subfloat[8192 processes]{\includegraphics[width=0.45\textwidth]{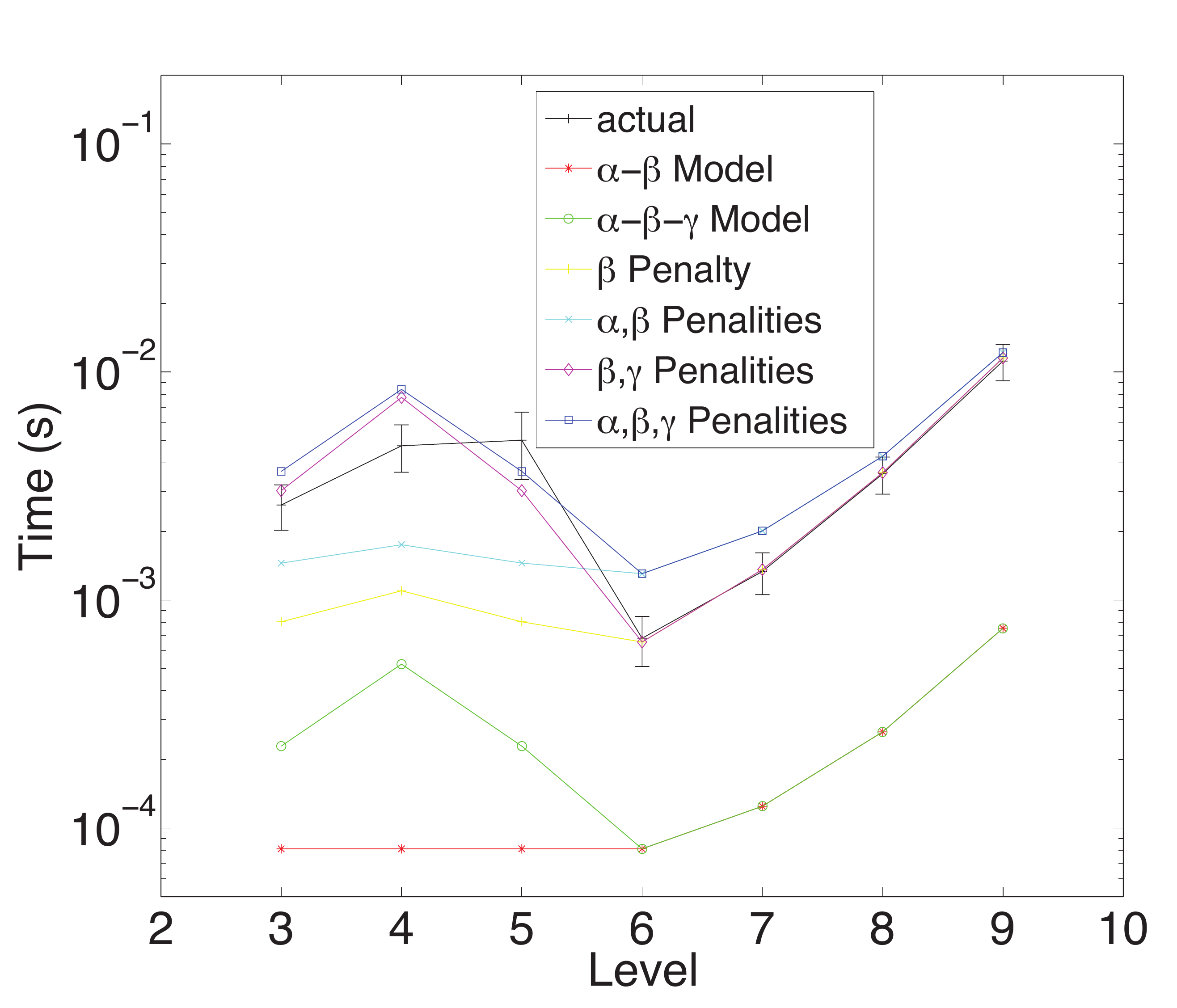}}
\caption{Performance model prediction and actual time for M2L communication phase on Titan.}
\label{fig:titan}
\end{figure}

Figure \ref{fig:titan} shows the M2L communication time on Titan along with the six model predictions. Similar to the previous two cases, the difference between the model predictions is mainly due to the correction for the inverse bandwidth. This difference in the theoretical injection bandwidth and measured effective bandwidth seems to have the largest effect on all three architectures. What is different from the previous two cases is the large jump in the actual communication time for the ``Global M2L" phase. For example, for the 8192 process run level 5 is taking about 10 times more than level 6 even though the message size is $46,592$ Bytes for both cases. The $\gamma$ term in the current performance models anticipates such behavior. The error bars in the actual timings are quite large, which indicates that there is a large load imbalance compared to the previous two systems.

\section{Conclusion}
The goal of this work is to model the global communication of the FMM and anticipate challenges on future exascale machines. To improve model fidelity, we consider penalties based on machine constraints including distance effects, reduced per core bandwidth, and the number of cores per node. We observe a good match between the \(\alpha - \beta - \gamma\) model with multicore penalties and the actual communication time. The discrepancy between the other models means that all components of the model; latency \(alpha\), bandwidth \(beta\), hops \(gamma\), and multicore penalty must be taken into account when predicting the communication performance of FMM.

In our benchmark tests, we compare the performance models with the actual measurements for the M2L communication, since this is the dominant part of the FMM communication. Our observations agree with that of the studies by Gahvari et al. \cite{Gahvari2011} where the performance of an algebraic multigrid method is analyzed using the same model. Our measurements fall within the bounds of the performance models, and match best with the model where latency, bandwidth, hops, and multicore penalty are all taken into account.

We were able to show that the present communication model is able to predict the performance on three HPC systems with different characteristics. To our knowledge, this is the first formal characterization of inter-node communication in FMM, which validates the model against actual measurements of communication time. We believe this is a step in the right direction, and the next logical step would be to increase the number of processes and continue to benchmark new HPC systems as they become available.

\section*{Acknowledgements}
The runs on Shaheen were provided through KAUST supercomputing lab. This research used resources of the Argonne Leadership Computing Facility at Argonne National Laboratory, which is supported by the Office of Science of the U.S. Department of Energy under contract DE-AC02-06CH11357. This research used resources of the Oak Ridge Leadership Computing Facility at the Oak Ridge National Laboratory, which is supported by the Office of Science of the U.S. Department of Energy under Contract No. DE-AC05-00OR22725.

\section*{Author Biographies}

\emph{Huda Ibeid} received her BSc degree in Computer Engineering from the University of Jordan and currently getting her PhD degree in Computer Science from King Abdullah University of science and Technology (KAUST). Her research interests include fast algorithms for particle-based simulations, fast algorithms on parallel computers and GPUs, design of parallel numerical algorithms, parallel programming models and performance optimizations for heterogeneous GPU-based systems. 

\bigskip

\noindent\emph{Rio Yokota} obtained his PhD from Keio University, Japan, in 2009 and went on to work as a postdoctoral researcher with Prof.\ Lorena Barba at the University of Bristol and then  Boston University.   He has worked on the implementation of fast $N$-body algorithms on special-purpose machines such as \textsc{mdgrape}-3, and then on GPUs after CUDA was released, and on vortex methods for fluids simulation. He joined the King Abdullah University of Science and Technology (KAUST) as a research scientist, where he continues to work on fast multipole methods.

\bigskip

\noindent\emph{David Keyes} is the director of the Strategic Initiative for Extreme Computing at KAUST and an affiliate of several laboratories of the U.S. Department of Energy (DOE). Keyes graduated in Aerospace and Mechanical Sciences from Princeton University and earned a doctorate in Applied Mathematics from Harvard University. He did postdoctoral work in the Computer Science Department of Yale University. With backgrounds in engineering, applied mathematics, and computer science, he works at the algorithmic interface between parallel computing and the numerical analysis of partial differential equations, across a spectrum of aerodynamic, geophysical, and chemically reacting flows.

\small
\bibliographystyle{plain}

\begin{thebibliography}{10}

\bibitem{Barnes1986}
J.~Barnes and P.~Hut.
\newblock {O($N \log N$)} force-calculation algorithm.
\newblock {\em Nature}, 324:446--449, 1986.

\bibitem{Beatson1997}
R.~Beatson and L.~Greengard.
\newblock A short course on fast multipole methods.
\newblock In {\em Wavelets, Multilevel Methods and Elliptic PDEs}, pages 1--37.
  Oxford Science Publications, 1997.

\bibitem{Vuduc2010}
A.~Chandramowlishwaran, S.~Williams, L.~Oliker, I.~Lashuk, G.~Biros, and
  R.~Vuduc.
\newblock Optimizing and tuning the fast multipole method for state-of-the-art
  multicore architectures.
\newblock In {\em Proceeding of the International Parallel Distributed
  Processing Symposium (IPDPS)}, pages 1--12, 2010.

\bibitem{Cheng1999}
H.~Cheng, L.~Greengard, and V.~Rokhlin.
\newblock A fast adaptive multipole algorithmin three dimensions.
\newblock {\em Journal of Computational Physics}, 155(2):468--498, 1999.

\bibitem{Clement1995}
M.~J. Clement and M.~J. Quinn.
\newblock Symbolic performance prediction of scalable parallel programs.
\newblock In {\em Proceedings of the International Parallel Processing
  Symposium}, pages 635--639, April 1995.

\bibitem{DeRose1999}
L.~DeRose and D.~A. Reed.
\newblock Svpablo: A multi-language, architecture-independent performance
  analysis system.
\newblock In {\em Proceeding of the International Conference on Parallel
  Processing}, pages 311--318, Augest 1999.

\bibitem{Dongarra2000}
J.~Dongarra and F.~Sullivan.
\newblock {Guest Editors Introduction to The Top 10 Algorithms}.
\newblock {\em Computing in Science and Engineering}, 2:22--23, 2000.

\bibitem{Foster1995}
I.~Foster.
\newblock {\em Designing and Building Parallel Programs}.
\newblock Addison-Wesley, 1995.

\bibitem{Foster1997}
I.~T. Foster and P.~H. Worley.
\newblock Parallel algorithms for the spectral transform method.
\newblock {\em SIAM Journal on Scientific and Statistical Computing},
  18(3):806--837, 1997.

\bibitem{Gahvari2011}
H.~Gahvari, A.~H. Baker, M.~Schulz, U.~M. Yang, K.~E. Jordan, and W.~Gropp.
\newblock Modeling the performance of an algebraic multigrid cycle on {HPC}
  platforms.
\newblock In {\em ICS '11 Proceedings of the International Conference on
  Supercomputing}, pages 172--181, 2011.

\bibitem{Gorn2004}
N.~L. Gorn and D.~V. Berkov.
\newblock Adaptation and performance of the fast multipole method for dipolar
  systems.
\newblock {\em Journal of Magnetism and Magnetic Materials}, 272-276:698--700,
  2004.

\bibitem{Greengard1996a}
L.~Greengard, M.~C. Kropinski, and A.~Mayo.
\newblock Integral equation methods for stokes flow and isotropic elasticity in
  the plane.
\newblock {\em Journal of Computational Physics}, 125:403--414, 1996.

\bibitem{Greengard1987}
L.~Greengard and V.~Rokhlin.
\newblock A fast algorithm for particle simulations.
\newblock {\em Journal of Computational Physics}, 73(2):325--348, 1987.

\bibitem{Greengard1988}
L.~Greengard and Rokhlin V.
\newblock On the efficient implementation of the fast multipole algorithm.
\newblock Research Report RR-602, Yale University, 1988.

\bibitem{Gropp1999}
W.~D. Gropp, D.K. Kaushik, D.E. Keyes, and B.F. Smith.
\newblock Toward realistic performance bounds for implicit {CFD} codes.
\newblock In {\em Proceedings of Parallel CFD'99}, pages 23--26, May 1999.

\bibitem{Jetley2010}
P.~Jetley, L.~Wesolowski, F.~Gioachin, L.~V. Kale, and T.~R. Quinn.
\newblock Scaling hierarchical {N}-body simulations on {GPU} clusters.
\newblock In {\em SC '10 Proceedings of the 2010 ACM/IEEE International
  Conference for High Performance Computing, Networking, Storage and Analysis},
  pages 1--11, 2010.

\bibitem{Kerbyson2001}
D.~Kerbyson, H.~Alme, A.~Hoisie, F.~Petrini, A.~Wasserman, and M.~Gittings.
\newblock Predictive performance and scalability modeling of a large-scale
  application.
\newblock In {\em Proceedings of the 2001 ACM/IEEE conference on
  Supercomputing}, pages 1--12, 2001.

\bibitem{Lashuk2009}
I.~Lashuk, A.~Chandramowlishwaran, H.~Langston, T.-A. Nguyen, R.~Sampath,
  A.~Shringarpure, R.~Vuduc, L.~Ying, D.~Zorin, and G.~Biros.
\newblock A massively parallel adaptive fast multipole method on heterogeneous
  architectures.
\newblock In {\em Proceedings of the Conference on High Performance Computing
  Networking, Storage and Analysis}, pages 1--12, 2009.

\bibitem{lee2013}
M.~Lee, N.~Malaya, and R.~D. Moser.
\newblock Petascale direct numerical simulation of turbulent channel flow on up
  to 768k cores.
\newblock In {\em Proceedings of the Conference on High Performance Computing
  Networking, Storage and Analysis}, Denver, CO, USA, Novermber 16-22 2013.

\bibitem{Luszczek2005}
P.~Luszczek and J.~Dongarra.
\newblock Introduction to the {HPCChallenge Benchmark Suite}.
\newblock {Technical Report ICL-UT-05-01}, university of Tennessee, Knoxville,
  March 2005.

\bibitem{Mendes1997}
C.~L. Mendes.
\newblock {\em Performance Scalability Prediction on Multicomputers}.
\newblock Phd thesis, University of Illinois, Urbana-Champaign, May 1997.

\bibitem{Mendes1998}
C.~L. Mendes and D.~A. Reed.
\newblock Integrated compilation and scalability analysis for parallel systems.
\newblock {\em International Conference on Parallel Architectures and
  Compilation Techniques (PACT'98)}, pages 385-- 392, October 1998.

\bibitem{Perez-Jorda1998}
J.~M. Perez-Jorda and W.~Yang.
\newblock On the scaling of multipole methods for particle-paticle
  interactions.
\newblock {\em Chemical Physics Letters}, 282:71--78, 1998.

\bibitem{Rankin1999}
W.~T. Rankin.
\newblock {\em Efficient Parallel Implementations of Multipole Based {N}-body
  Algorithm}.
\newblock PhD thesis, Duke University, 1999.

\bibitem{Snavely2001}
A.~Snavely, N.~Wolter, and L.~Carrington.
\newblock Modeling application performance by convolving machine signatures
  with application profiles.
\newblock In {\em Proceeding of the IEEE Workshop on Workload
  Characterization}, pages 149--156, December 2001.

\bibitem{VandeWiele2008}
B.~Van~de Wiele, F.~Olyslager, and L.~Dupre.
\newblock Application of the fast multipole method for the evaluation of
  magneto-static fields in micromagnetic computations.
\newblock {\em Journal of Computational Physics}, 227:9913--9932, 2008.

\bibitem{Wolf2011}
W.~R. Wolf and S.~K. Lele.
\newblock Aeroacoustic integrals accelerated by fast multipole method.
\newblock {\em AIAA Journal}, 49(7):1466--1477, 2011.

\bibitem{Worley2000}
P.~H. Worley.
\newblock Performance evaluation of the {IBM SP and the Compaq AlphaServer SC}.
\newblock In {\em Proceeding of the ACM International Conference of
  Supercomputing 2000}, pages 235--244, 2000.

\bibitem{Zhao2000}
J.-S. Zhao and W.-C. Chew.
\newblock Three-dimensional multilevel fast multipole algorithm from static to
  electrodynamic.
\newblock {\em Microwave and Optical Technology Letters}, 26(1):43--48, 2000.

\end{thebibliography}

\end{document}